# TapToTab : Video-Based Guitar Tabs Generation using AI and Audio Analysis

Ali Ghaleb
*Faculty of Computer and Information Sciences*
*Ain Shams University*
*Cairo, Egypt*
20201701633@cis.asu.edu.eg

Eslam Elsadawy
*Faculty of Computer and Information Sciences*
*Ain Shams University*
*Cairo, Egypt*
20201701706@cis.asu.edu.eg

Ihab Essam
*Faculty of Computer and Information Sciences*
*Ain Shams University*
*Cairo, Egypt*
20201701710@cis.asu.edu.eg

Seif-Eldin Zaky
*Faculty of Computer and Information Sciences*
*Ain Shams University*
*Cairo, Egypt*
20201701718@cis.asu.edu.eg

Mohamed Abdelhakim
*Faculty of Computer and Information Sciences*
*Ain Shams University*
*Cairo, Egypt*
20201701735@cis.asu.edu.eg

Natalie Fahim
*Faculty of Computer and Information Sciences*
*Ain Shams University*
*Cairo, Egypt*
20201701743@cis.asu.edu.eg

Razan Bayoumi
*Faculty of Computer and Information Sciences*
*Ain Shams University*
*Cairo, Egypt*
razan.bayoumi@cis.asu.edu.eg

Hanan Hindy
*Faculty of Computer and Information Sciences*
*Ain Shams University*
*Cairo, Egypt*
hanan.hindy@cis.asu.edu.eg

*Abstract*— The automation of guitar tablature generation from video inputs holds significant promise for enhancing music education, transcription accuracy, and performance analysis. Existing methods face challenges with consistency and completeness, particularly in detecting fretboards and accurately identifying notes. To address these issues, this paper introduces an advanced approach leveraging deep learning, specifically YOLO models for real-time fretboard detection, and Fourier Transform-based audio analysis for precise note identification. Experimental results demonstrate substantial improvements in detection accuracy and robustness compared to traditional techniques. This paper outlines the development, implementation, and evaluation of these methodologies, aiming to revolutionize guitar instruction by automating the creation of guitar tabs from video recordings.

*Keywords— Automated Guitar Transcription, Computer Vision, YOLO, Deep Learning, Frequency Analysis, Fourier Transform, Image Processing, Canny Edge Detection, Fretboard Detection, Audio-visual Integration, Guitar Tablature Generation, Real-time Performance Analysis.*

## I. Introduction

The accurate generation of guitar tablature is a significant challenge in the fields of music analysis, education, and transcription. Traditional methods that rely solely on audio signal processing often struggle with accuracy due to the inherent complexities of audio signals and the physical layout of the guitar fretboard [1]. The variations in appearance, texture, and lighting conditions of different guitars further complicate the task when using conventional image processing techniques like Canny edge detection. This paper presents an innovative approach that leverages the advancements in deep learning and computer vision, combined with robust audio analysis, to develop a model capable of real-time guitar tablature generation from video inputs.

In recent years, Convolutional Neural Networks (CNNs) have revolutionized image recognition tasks, providing unprecedented accuracy in object detection and classification [2]. Building upon this foundation, our paper employs You Only Look Once (YOLO) models, specifically YOLO V8 and YOLO V8 Oriented Bounding Box (OBB), to detect and segment guitar strings and fret zones with high precision. YOLO models are chosen for their real-time performance, efficiency, and robustness against occlusions and overlapping objects, making them ideal for this application.

Simultaneously, the audio analysis component of our paper utilizes Fourier Transform to convert audio signals from the time domain to the frequency domain, enabling precise identification of note frequencies. By integrating these visual and auditory data streams, our paper provides a comprehensive solution that surpasses the accuracy of existing audio-only approaches.

The primary objective of this paper is to propose a complete model that can automatically generate guitar tablature by accurately detecting the fretboard and identifying played notes in real time. The results demonstrate significant improvements in the accuracy and robustness of fret line and note detection compared to traditional methods reaching YOLO V8 OBB, in particular, exhibits the highest precision and recall metrics, establishing it as the optimal choice for fretboard detection. The Fourier Transform-based frequency analysis effectively identifies guitar notes from video audio, contributing to the overall accuracy and efficiency of the system.

This paper outlines the development and implementation of the proposed models, including the methodologies used for video and audio analysis, the integration of these data streams, and the experimental results. Through this interdisciplinary approach, we aim to provide a reliable and accurate tool for guitar tab generation, enhancing the learning and transcription experience for musicians. This could also improve way people learn specially after the COVID-19 pandemic [3].

The main contributions of this papers can be summarized as follows:

- Development of a novel system combining computer vision and audio analysis for accurate and real-time guitar tablature generation.
- Creation and augmentation of a specialized dataset for training and validating the proposed models, addressing the limitations of existing datasets.

The rest of this paper is organized as follows; Section II summarizes the prominent papers in the literature that aimed to solve this problem. The proposed methodology is discussed in Section III, followed by the dataset generation in Section IV. The implementation and results are outlined in Section V. Finally, the paper is concluded in Section VI.

## II. RELATED WORK

This section explores the commonly used approaches in the field of guitar tab generation, specifically focusing on computer vision and audio analysis methodologies. The following subsections provide a detailed discussion of these approaches.

### A. Computer Vision Techniques in Guitar Tab Generation

The authors in [4] describes the conventional approach to guitar fingering retrieval using Classical Computer Vision [5]. Their method involves two main tasks: feature extraction and classification. The paper discusses several feature extraction techniques, including background subtraction, Canny edge detection, probabilistic Hough transform, and horizontal Sobel filter [4]. These techniques are used to locate the fretboard and detect individual fretting fingers. The system proposed in the paper improves upon existing methods by eliminating the need for a fixed camera on the guitar neck and markers, achieving better accuracy in note recognition. The classification of the detected features is then performed to identify the positions of the guitarist's fingers. Preliminary testing shows that this markerless approach provides more accurate results compared to previous methods, as evidenced by the results in [4].

The proposed system in [6] is vision-based and markerless. It tracks the guitar's strings and frets and uses skin detection to localize the guitarist's fingers on the fretboard. This information is then used to generate tablature notation, indicating which strings and frets should be played for each beat. Key features of the system include Markerless Tracking, Real-time Processing, Fretboard Detection, Finger Detection.

### B. Audio Analysis Methods in Guitar Tab Generation

Kazuki Yazawa *et al.* [7] propose a sophisticated method for transcribing guitar tablature from audio signals. This method addresses the common challenge of estimating multiple pitches simultaneously while ensuring the resulting configurations are physically playable on a guitar. The proposed system integrates multipitch estimation and fingering configuration estimation, incorporating three critical constraints to eliminate unplayable combinations of pitches. The approach includes the following key elements:

- The system begins with multipitch estimation using Latent Harmonic Allocation (LHA), a machine learning method that estimates multiple pitches from observed spectrograms. The output indicates the likelihood of each pitch in each time frame, The method uses dynamic programming to optimize fingering configurations across time frames, ensuring temporal continuity and playability.
- The system enumerates all possible fingering configurations that are physically playable, constrained by the reach and number of fingers. Configurations are categorized into open chords and barre chords, with specific rules governing each type.

Experiments conducted with synthesized guitar sounds from MIDI data [7] demonstrated a significant improvement in multipitch estimation accuracy, with a 5.9-point increase in F-measure compared to conventional methods. The transcribed tablatures were found to be playable and more accurate.

Recent advancements in AI have significantly improved the accuracy and efficiency of generating guitar tablature from audio and video inputs. Various methodologies, including computer vision, deep learning, and machine learning techniques, have been employed to tackle the challenges associated with automatic guitar tab generation. These approaches have demonstrated notable improvements in note recognition, accuracy, and the ability to handle complex playing styles and fast note transitions.

Table 1 summarizes the results of recent papers that have contributed to the field of guitar tab generation using AI, highlighting their methodologies and key findings.

*Table 1. Results of the Recent Papers*

| Paper | Methodology | Key Findings |
|---|---|---|
| Retrieval of Guitarist Fingering Information using Computer Vision Feature Extraction and Classification [4] | Classical Computer Vision techniques: feature extraction and classification | Improved note recognition accuracy by eliminating the need for a fixed camera on the guitar neck and markers. |
| Guitar Tablature Generating Using Computer Vision [6] | Vision-based, markerless tracking, real-time processing | Achieved significantly higher accuracy in note identification and real-time tablature generation. |
| Audio-based guitar tablature transcription[7] | Multipitch estimation, fingering configuration estimation, machine learning | Demonstrated a significant improvement in multipitch estimation accuracy, transcribed playable and accurate tablatures. |
| Guitar Tablature Estimation with a Convolutional Neural Network [8] | Convolutional Neural Network (CNN) | Achieved high accuracy in guitar tablature estimation by learning complex patterns in audio spectrograms. |

## III. METHODOLOGY

This section highlights the main process of the proposed Guitar Tabs Generation method. The section is divided into Data Collection, Model Development, Integration

### A. Data Collection

While developing our automated guitar transcription system, the selection and utilization of appropriate datasets were crucial to our success. Initially, we explored several existing datasets to determine their suitability for our specific needs, detailed discussion of these approaches is outlined in Section IV.

*1) Video Data Collection*

To develop a robust system for guitar tablature generation, we collected video data of guitar performances. The video data captures the fretboard and the guitarist's hand movements, which are essential for accurate detection and analysis. The data has been collected from self-recorded videos as well as popular Youtube videos. The annotations can be requested directly from the authors .

*a) Recording Sessions: M*ultiple recording sessions were conducted featuring various guitarists playing different musical pieces on both acoustic and electric guitars. These sessions were designed to capture a wide range of playing styles, techniques, and fret positions.

*b) Equipment Used:* High-definition cameras were used to ensure clear visibility of the fretboard and hand movements. The cameras were positioned to provide an unobstructed view of the guitarist's hands and the guitar neck.

*2) Audio Data Collection*

In addition to the video data, audio data was meticulously collected to ensure accurate note identification. Audio data has been published and is publicly available at [9]

*3) Data Annotation*

The collected data was annotated using specialized tools to provide labels for training the machine learning models.

*a) Roboflow:* Roboflow was used to annotate the video data *Fig. 1*, segmenting it into 12 fret classes. This segmentation allowed us to accurately track the guitarist's hand positions across the fretboard, providing critical data for our computer vision algorithms.

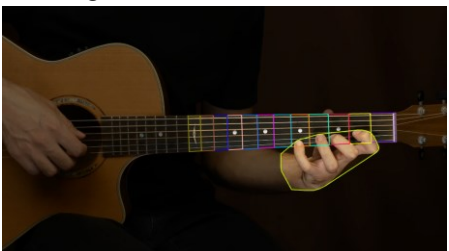

*Fig. 1. Roboflow Annotation*

### B. Model Development

*1) Video Data Collection*

We employed advanced computer vision models to detect and segment the guitar fretboard and the guitarist's hand movements.

YOLO Models: YOLO V8 and YOLO V8 OBB models were used for their real-time performance and high accuracy in detecting and segmenting objects.

*a) Training:* The YOLO models were trained on the annotated video data to detect the fretboard, strings, and hand positions. The training process involved optimizing the models to handle various lighting conditions, angles, and occlusions.

*b) Evaluation:* The performance of the models was evaluated using precision, recall, and Mean Average Precision (MAP) metrics to ensure high accuracy in object detection and segmentation.

*2) Audio Analysis*

For audio analysis, traditional signal processing techniques were used to identify the notes played.

*a) Frequency Analysis (FourierTransform):* Fourier Transform was used to convert audio signals from the time domain to the frequency domain, enabling precise identification of note frequencies, The audio signals were analyzed to detect the fundamental frequencies of the notes played. The identified frequencies were then mapped to corresponding guitar notes.

### C. Integration of Audio and Visual Data

To generate accurate guitar tablature, the audio and visual data analysis were integrated.

- Synchronization: The audio and video data were synchronized using onset detection as shown in *"Fig. 2"*, which identifies the precise moments when notes are played.

- Correlation**:** The visual data (hand positions on the fretboard) and the audio data (identified note frequencies) were correlated to determine the exact notes being played at each moment.

- Tablature Generation**:** Based on the correlated data, the system generated guitar tablature, indicating which strings and frets were played for each note.

By employing a combination of advanced computer vision and audio analysis techniques, our methodology ensures high accuracy and reliability in generating guitar tablature from video inputs. The integration of these techniques provides a comprehensive solution that overcomes the limitations of existing methods and offers a powerful tool for musicians and educators.

| Frame Nu | Onset Tim | Top Notes |
|---|---|---|
| 57 | 2.2639455 | [[332.0, 'E4' |
| 71 | 2.8212244 | [[332.0, 'E4' |
| 72 | 2.8792743 | [[332.0, 'E4' |
| 86 | 3.4133333 | [[332.0, 'E4' |
| 88 | 3.4946031 | [[332.0, 'E4' |
| 103 | 4.0983219 | [[332.0, 'E4' |
| 116 | 4.6207709 | [[332.0, 'E4' |

*Fig. 2. Onset Time corresponding to Frame number and Top Note*

## IV. DATASET

While developing our automated guitar transcription system, the selection and utilization of appropriate datasets were crucial to our success. Initially, several existing datasets were explored to determine their suitability for our specific needs.

*1) Existing Datasets*
- Synch GuitarSet [10]**:** The GuitarSet dataset provides high-quality guitar recordings along with extensive annotations and metadata. This dataset includes recordings of various musical excerpts played on an acoustic guitar *"Fig.3"*, with time-aligned annotations that detail pitch contours, string and fret positions, chords, beats, downbeats, and playing style. While this dataset has rich annotations and high quality, it primarily focuses on acoustic guitars and may not fully address the needs of electric guitar transcription or provide the necessary visual data for our computer vision approach.

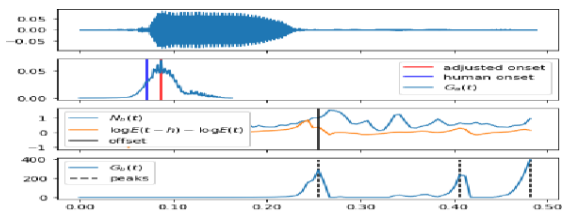

*Fig. 3. GuitarSet Dataset [10]*

- IDMT-SMT-GUITAR [11]: This large database is designed for automatic guitar transcription and features recordings from seven different guitars in standard tuning, with varying pick-up settings and string measures. While the IDMT-SMT-GUITAR dataset offers a diversified collection of both electric and acoustic guitars, it predominantly focuses on audio data. Our model requires a dataset that includes comprehensive visual data to enable effective computer vision analysis.

- DAVIS [12]: The DAVIS dataset is a high-quality and high-resolution video segmentation dataset, offering dense annotations at resolutions of 480p and 1080p. While it provides excellent video data for segmentation tasks [13], it is not specifically tailored to musical instruments or guitar playing, lacking the musical annotations and guitar-specific data required for our project.

While these datasets are invaluable resources, they each have limitations that impact their effectiveness for our specific scope:

- Lack of Visual Guitar Data: GuitarSet and IDMT-SMT-GUITAR primarily focus on audio data with limited visual information, which is essential for our computer vision approach.

- Specificity to Acoustic Guitars: GuitarSet focuses on acoustic guitars, while our project aims to include both acoustic and electric guitars.

- Generic Segmentation Data: DAVIS provides excellent video segmentation data but is not tailored to guitar playing, lacking necessary musical annotations and context.

To address these challenges, we undertook the task of self-collecting a dataset using Roboflow, a tool designed to facilitate the creation and management of computer vision datasets. Our goal was to create a dataset that provided both high-quality audio and visual data, specifically tailored to guitar playing following two solutions. The experiments in which these datasets were utilized is discussed in detail in Section V.

- Zone Annotation Approach: Annotate frets to map specific zones on the fretboard for efficient hand detection used in (Experiment #3).

- Fretboard Segmentation Approach: Segment the fretboard into distinct areas for precise finger placement monitoring used in (Experiment #4).

Despite these challenges, the effort invested in self-collecting and annotating our dataset was essential for the success of our project. The resulting dataset provided a robust foundation for developing our automated guitar transcription system, combining the strengths of existing datasets with customized data tailored to our specific needs.

By leveraging advanced tools like Roboflow and committing to meticulous data collection and annotation processes, we were able to overcome the limitations of existing datasets and create a comprehensive resource that significantly enhanced the accuracy and usability of our transcription system.

To further enhance the dataset's effectiveness, various augmentation techniques were applied to increase the diversity and volume of our training data table as shown in Table 2. Augmentation helps in improving the robustness of the model by providing it with varied training examples.

*Table 2. Generated Dataset Volume*

| Approach | Before Augmentation | After Augmentation | Augmentation Techniques |
|---|---|---|---|
| **Zone Annotation Approach** | 440 | 1100 | Flip, Shear, Noise |
| **Fretboard Segmentation Approach** | 730 | 1790 | Flip, Rotate, Saturation, Exposure, Mosaic |

In addition to video data, audio data was meticulously collected to ensure accurate note identification. The dataset is entitled "TapToTab: A Pitch-Labelled Guitar Dataset for Note Recognition" and is available at [9].

  *a) Recording Process:* A guitarist played all notes up to the 12th fret on all six strings, both in clean and distorted settings. This was done to capture the full range of notes and tones produced by the guitar.

  *b) Manual Annotation:* Each audio sample was manually annotated with the note name and its pitch (e.g., E4). This detailed annotation was crucial for training the audio analysis component of the system.

  *c) Data Augmentation:* To enhance the dataset's diversity and size, various augmentation techniques were applied to the audio samples. This included adding noise, shifting pitch, and altering the speed of the recordings.

## V. IMPLEMENTATION, EVALUATION AND RESULTS

The section encompasses a series of experiments conducted to evaluate and refine our automated guitar transcription system. Each experiment aimed to address specific challenges in detecting and segmenting guitar fretboards, as well as, accurately identifying the notes played. By employing a combination of image processing techniques and advanced deep learning models, we sought to enhance the precision and robustness of our system.

### A. Video Detection Experiments

*1) Experiment #1 (Image Processing)*

In this experiment, the objective was to explore the feasibility of automatically generating guitar tabs from video footage using specialized techniques. The primary focus was on identifying the vibrating guitar strings through a sequential workflow of image processing stages. Initially, the video frames were converted to grayscale to simplify subsequent analyses, followed by resizing to ensure uniform resolution. The core techniques employed included contour detection and the Hough Line Transform as shown in Fig. 4.

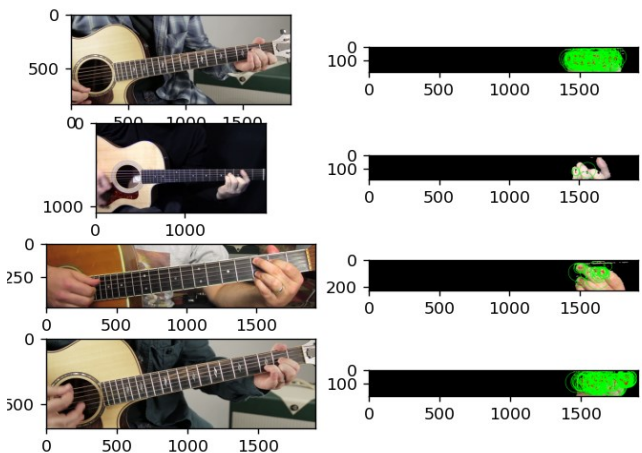

*Fig. 4. Circular Hough Transform*

Manual Adjustment of the Region of Interest (ROI) requires manual setting of the ROI for accurate analysis, which introduces inefficiencies. Constant adjustments may be needed due to variations in lighting conditions, camera angles, and other environmental factors. This manual intervention contradicts the goal of creating an automated and user-friendly system, limiting scalability and usability.

*2) Experiment #2 (Canny Edge Detection)*

The experiment involves applying the Canny edge detection algorithm, a multi-step process designed to detect a wide range of edges in images.

When applied to the preprocessed video frames of a guitar fretboard, the Canny edge detection algorithm highlights potential fret lines by identifying the edges corresponding to the boundaries of the frets, shown in *Fig. 5*. However, due to variations in appearance, texture, and lighting conditions of different guitar fretboards, the algorithm may produce inconsistent results, leading to the rejection of this technique for reliable fret line detection.

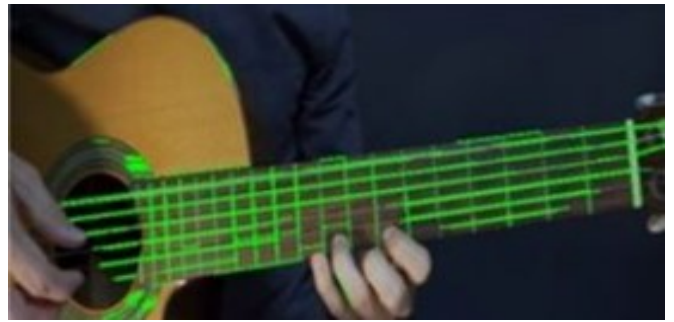

*Fig. 5. Canny Edge Detection*

Rejection of Canny Edge Detection: After thorough testing, it was concluded that the Canny edge detection technique is not suitable for our purpose. The variability in fretboard appearances and lighting conditions resulted in inconsistent detection of fret lines.

To enhance the accuracy and robustness of fret line detection, we propose transitioning to deep learning techniques. Deep learning models can learn to recognize fret lines across various conditions by training on a diverse dataset of fretboard images.

*3) Experiment #3 (YOLO Models)*

The experiment involves applying a fret detection algorithm to identify guitar strings and individual zones. This method aims to provide precise segmentation of the guitar fretboard, allowing for accurate detection of fret lines and notes.

In addition to detecting each fret as a zone, the experiment also includes the detection of the left hand and computing the Intersection over Union (IoU) for each fret and the left hand. This approach allows us to predict which fret has been played based on the highest overlap between the left hand and the detected zones.

The experiment demonstrates that deep learning techniques, particularly using YOLO models, significantly improve the accuracy and robustness of fret line detection compared to traditional methods like the Canny edge detection algorithm. YOLO V8 and its OBB variant show the highest precision and recall metrics, indicating superior performance in detecting and segmenting guitar fretboards.

YOLO V8 OBB's ability to account for object orientation leads to the most accurate results, making it the best choice for our application. The results of the different experiments conducted using YOLO is shown in Table 2, and the validation batch prediction shown in *Fig.6*.

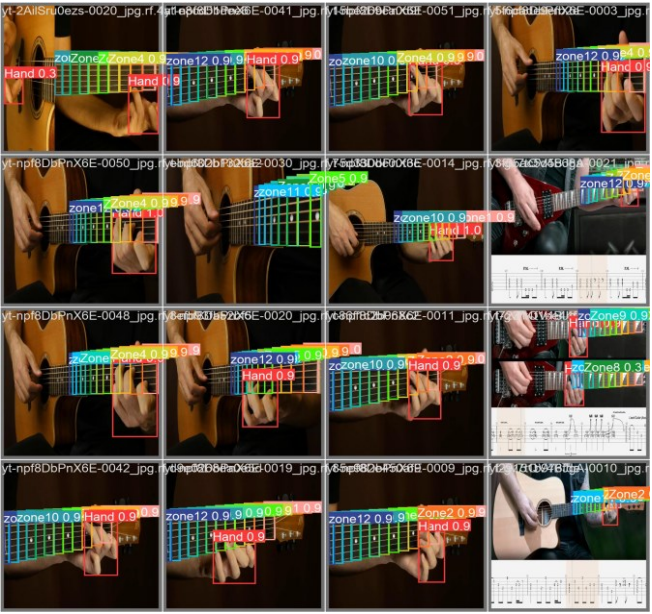

*Fig. 6. Experiment 3 Results*

*Table 3. YOLO Versions Results and Metrics*

| Model | Precision | Recall | MAP50 | MAP50-95 |
|---|---|---|---|---|
| **YOLO V5** | 0.919 | 0.822 | 0.913 | 0.769 |
| **YOLO V8** | 0.987 | 0.995 | 0.992 | 0.951 |
| **YOLO V8 OBB** | 0.988 | 0.988 | 0.993 | 0.955 |
| **YOLO V9** | 0.958 | 0.949 | 0.983 | 0.907 |

*4) Experiment #4 (Advanced YOLO and Zone Division)*

The primary goal of this experiment is to develop a deep learning algorithm for detecting and segmenting the guitar fretboard. The algorithm is designed to accurately identify the strings and individual fret zones, enhancing the precision of note detection during guitar performances.

Based on the provided frequency chart in *Fig.7*, the same note frequency can be played on three different frets, spaced equally apart. This simplifies the detection process by reducing the number of fret positions that need to be distinguished.

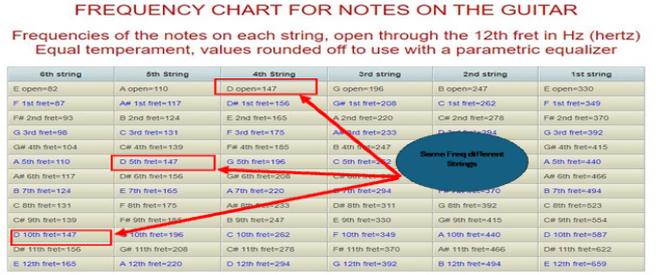

*Fig. 7. Frequency Chart*

By implementing this technique, the number of classes is significantly reduced to two

- Left Hand Position Segment 1
- Fretboard

Each zone is treated as a distinct class or category within our system. For instance, Zone 1 might encompass the first segment of the fretboard (e.g., frets 0 to 4 on an acoustic guitar), while Zone 2 could cover the subsequent segment (e.g., frets 5 to 9), and so forth as shown in (1).

This classification allows our system to categorize the location of the guitarist's hand based on the detected position relative to these predefined zones. By associating each frame's detected hand position with a specific zone, our system can accurately determine the fret range where notes are being played.

$$ZONE = f / (max\_frets/num\_zones) \quad (1)$$

This equation dynamically assigns each fret position $f$ to the appropriate zone based on the guitar type, ensuring accurate localization and transcription of played notes.

*B. Audio Analysis Trials*

*1) Experiment #1 (Frequency Analysis)*

A frequency analysis algorithm to extract the frequency of each note from the guitar audio. By mapping these frequencies to a predefined list of guitar note frequencies *"Fig.7"*, we can accurately identify and label each note.

The experiment demonstrates that frequency analysis using Fourier Transform provides an accurate and efficient method for detecting and identifying guitar notes from video audio. The approach outperforms traditional methods by leveraging precise frequency resolution and efficient processing capabilities.

*2) Experiment #2 (Deep Learning)*

A deep learning model to analyze the audio extracted from a guitar video. The model is trained on a guitar notes dataset available on Kaggle [14, 15] to recognize different guitar notes, enabling precise note identification.

The experiment demonstrates that deep learning techniques provide an accurate and efficient method for detecting and identifying guitar notes from video audio. The approach leverages the powerful pattern recognition capabilities of deep learning models to achieve high accuracy in note identification.

## C. Note Mapping Process

A crucial step that bridges the audio and visual data to produce accurate tablature, Accurate note mapping is essential for ensuring that the transcriptions reflect the precise timing and positions of the notes played during a guitar performance. By leveraging both audio analysis and computer vision techniques, we can achieve a detailed and synchronized understanding of the music being played.

- Video and Audio Acquisition: Initially, we acquire a video of the guitar performance, which includes both visual and audio data. This combined data serves as the foundation for our subsequent analysis.

- Audio Analysis and Onset Detection: We begin with audio analysis, focusing on onset detection. Onset detection identifies the precise moments when notes are played, capturing both the frequency and the exact timing of each note. This step is essential as it allows us to correlate specific audio events with visual data.

- Synchronizing Video Frames with Audio Onsets: Using the timestamps from the onset detection, we segment the video into frames that correspond to the detected audio events. This synchronization ensures that each video frame analyzed is directly linked to a specific note onset, providing a temporal connection between the audio and visual data.

- Computer Vision and Inference: We then apply computer vision techniques to these synchronized video frames. Our model segments the guitar fretboard into 12 distinct zones, each representing a fret. By analyzing the frames, we detect the positions of the player's hands and determine which zones on the fretboard they overlap with at the exact moments of note onset.

- Mapping Zones to Notes: Finally, we combine the information from the audio (frequencies and onsets) with the visual data (detected hand positions on the fretboard). By correlating the detected zones on the fretboard with the frequencies from the audio analysis, we accurately identify which fret and string were played for each note.

*Fig.8*, a form of musical notation indicating instrument fingering rather than musical pitches. The tablature consists of six lines representing the six strings of the guitar. Numbers on the lines indicate which fret to press down on the respective string.

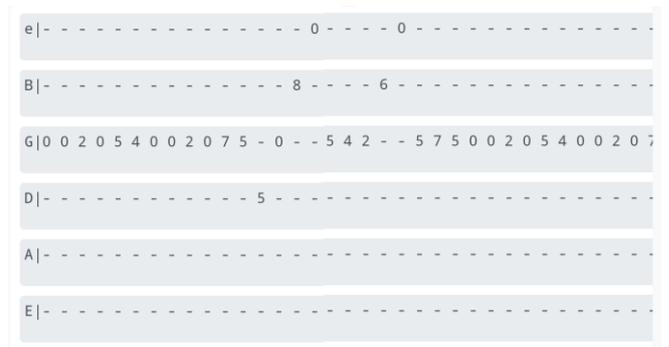

*Fig. 8. Guitar Tablature Notation Result*

This integrated approach of audio and visual analysis allows us to generate precise guitar tabs, reflecting both the temporal and spatial aspects of the performance. Our method ensures a high level of accuracy in detecting and mapping notes, providing a reliable tool for guitarists seeking to transcribe music from video performances.

## VI. Conclusion and Future work

This paper presents a comprehensive system for generating guitar tablature from video inputs by leveraging advanced computer vision and audio analysis techniques. Traditional methods for guitar tab generation, such as Canny edge detection, have shown significant limitations due to variations in the appearance, texture, and lighting conditions of guitar fretboards. Our proposed system overcomes these challenges by employing deep learning models, specifically YOLOv8 and YOLOv8 OBB, for precise fretboard detection and segmentation, along with Fourier Transform-based algorithms for accurate audio analysis.

The integration of these methodologies enables real-time detection and segmentation of guitar fretboards, leading to precise note identification and automatic tablature generation. Experimental results demonstrate that our approach significantly improves accuracy and robustness compared to traditional methods, providing a reliable solution for musicians and educators. This system not only enhances the learning experience for guitarists by automating the tab generation process but also contributes to the field of music education technology.

While the current models show promising results, there are several areas for future improvement and expansion:

*1) Handling Extreme Occlusions and Rapid Note Changes:* Future research should focus on refining the models to handle extreme cases of occlusion and rapid note transitions more effectively. This could involve developing more sophisticated algorithms for tracking hand movements and note transitions in complex scenarios.

*2) Expanding the Dataset:* To improve the robustness and generalizability of the system, it is essential to expand the dataset to include a wider variety of guitars, playing styles, and environmental conditions. A more diverse dataset will ensure that the model can handle different scenarios and provide accurate results across various contexts.

*3)* Integration into Real-Time Applications: One of the most promising directions for future work is integrating the detection algorithm into a real-time application. Developing a mobile app or desktop software that provides live performance analysis and feedback to guitarists can enhance their learning experience and enable immediate improvements.

*4)* Improving Model Efficiency: While YOLO models are known for their efficiency, further optimizations can be made to reduce computational requirements and improve processing speed. This is particularly important for real-time applications where low latency is crucial.

*5)* User Interface and Usability Enhancements: Enhancing the user interface and overall usability of the application will make it more accessible to a broader audience, including beginners and those with limited technical expertise. User feedback should be incorporated to continuously improve the design and functionality of the system.

*6)* Incorporating Advanced Audio Processing Techniques: Future iterations of the system can explore more advanced audio processing techniques, such as machine learning-based pitch detection algorithms, to further enhance the accuracy of note identification.

*7)* Multi-Instrument Support: Extending the system to support other string instruments, such as bass guitar, violin, or ukulele, could broaden its applicability and appeal to a wider range of musicians.

REFERENCES

[1] E. B. Ayyıldız and O. Zahal, "Instructor experiences with online guitar lessons during the Covid-19 pandemic in Turkey," International Journal of Music Education, vol. 41, no. 3, pp. 484-496, 2023.

[2] A. Wiggins and Y. E. Kim, "Guitar Tablature Estimation with a Convolutional Neural Network," Proceedings of the 20th ISMIR Conference, Delft, Netherlands, pp. 284-291, 2019.

[3] K. Kaspar, K. Burtniak and M. Rüth, "Online learning during the Covid-19 pandemic: how university students' perceptions, engagement, and performance are related to their personal characteristics," vol. 43, no. 18, pp. 16711-16730, 2024.

[4] R. G. Joseph Scarr, "Retrieval of guitarist fingering information using computer vision," in 2010 25th International Conference of Image and Vision Computing New Zealand, 2010.

[5] "What is computer vision?," IBM, [Online]. Available: https://www.ibm.com/topics/computer-vision. [Accessed 21 6 2024].

[6] B. Duke and A. Salgian, "Guitar Tablature Generation using Computer Vision," Advances in Visual Computing: 14th International Symposium on Visual Computing, ISVC 2019, Lake Tahoe, NV, USA, October 7--9, 2019, Proceedings, Part II 14, pp. 247-257, 2019.

[7] K. Yazawa, D. Sakaue, K. Nagira, K. Itoyama and H. G. Okuno, "Audio-based guitar tablature transcription using multipitch analysis and playability constraints," IEEE International Conference on Acoustics, Speech and Signal Processing, pp. 196-200, 2013.

[8] R. Vanegas, "The MIDI pick: Trigger serial data, samples, and MIDI from a guitar pick," in Proceedings of the 7th international conference on New interfaces for musical expression, 2007.

[9] E. ElSadawy, I. Essam, A. Ghaleb, M. Abdelhakim, S. Zaki, N. Fahim, R. Bayoumi, H. Hindy, September 4, 2024, "TapToTab: A Pitch-Labelled Guitar Dataset for Note Recognition", IEEE Dataport, doi: https://dx.doi.org/10.21227/664p-5b45.

[10] X. Wen, R. M. Bittner, J. Salamon and J. P. Bello, "GuitarSet: A Dataset for Guitar Transcription," 2018. [Online]. Available: https://guitarset.weebly.com/. [Accessed 29 6 2024].

[11] C. Kehling, A. Männchen and A. Eppler, "IDMT-SMT-Guitar Dataset," [Online]. Available: https://www.idmt.fraunhofer.de/en/publications/datasets/guitar.html. [Accessed 29 6 2024].

[12] "DAVIS Challenge," DAVIS, 2024. [Online]. Available: https://davischallenge.org/. [Accessed: Jun. 29, 2024].

[13] C. Pan, Y. Yang, K. Mo, Y. Duan and L. Guibas, "Object Pursuit: Building a Space of Objects via Discriminative Weight Generation," arXiv preprint, 2021.

[14] W. H. Elashmawi, J. Emad, A. Serag, K. Khaled, A. Yehia, K. Mohamed, H. Sobeah and A. Ali, "A Novel Approach for Improving Guitarists' Performance Using Motion Capture and Note Frequency Recognition," Applied Sciences, vol. 13, no. 10, p. 6302, 2023.

[15] Y. Aznalhakiky, "Guitar Notes," Kaggle, 2023. [Online]. Available: https://www.kaggle.com/datasets/yulianaznalhakiky/guitar-notes. [Accessed: Jun. 29, 2024].